\documentclass[12pt,a4paper]{article}
\usepackage{amssymb}

\usepackage{latexsym}
\usepackage{amsmath}

\newtheorem{theorem}{Theorem}

\newtheorem{remark}[theorem]{Remark}

\numberwithin{equation}{section}

\begin{document}

\title{A shortcut to sign Incremental Value-at-Risk for risk allocation\footnote{%
The opinions expressed in this paper are those of the authors and do not
necessarily reflect opinions shared by Deutsche Bundesbank.} }
\author{Dirk Tasche\thanks{
Deutsche Bundesbank, Postfach 10 06 02, 60006 Frankfurt am Main, Germany;
e-mail:\newline tasche@ma.tum.de} \quad Luisa Tibiletti\thanks{%
Corresponding Author: Luisa Tibiletti, Associate Professor of Mathematics
for Economics and Finance, Dipartimento di Statistica e Matematica, Universit%
\`{a} di Torino, Piazza Arbarello, 8, 10122 Torino, Italy; fax
++39-11-6706238; e-mail: luisa.tibiletti@unito.it}}
\date{October 2, 2002}
\maketitle

\begin{abstract}
%% {\sc AMS subject classification:} \textit{62G35; 62J05.}
Approximate Incremental Value-at-Risk formulae provide an easy-to-use
preliminary guideline for risk allocation. Both the cases of risk adding and
risk pooling are examined and beta-based formulae achieved. Results
highlight how much the conditions for adding new risky positions are
stronger than those required for risk pooling.


\end{abstract}

\pagebreak

\section{Introduction}

Incremental Value-at-Risk (IVaR) is becoming a standard tool in investment
management industry to identify strategies that enhance returns and control
risk. In theory, IVaR is a metrics that measures the contribution in terms
of relative Value-at-Risk (VaR) of a position or a group of positions with
respect to the total risk of a pre-existent portfolio. In the academic
literature the attention to this technique can be traced back to the some
works by Dowd [1998, 1999, 2000]. Nevertheless, in recent times more
practice-oriented researchers have dwelt on this tool for a twofold purpose:
1) hedging and speculating with options (see Mina [2002]) and 2) reducing
the risk in a risk-return analysis. Two slightly different approaches
follow: one leads to a risk adding model, the other to a risk pooling model.

Risk adding means increasing the amount of invested money by purchasing
a new asset. This new asset will be a hedge if the portfolio risk is reduced
by its addition. When risk is measured with VaR, risk reduction is indicated
by a negative IVaR.

Pooling risks means to keep the amount of invested money constant. Hence,
pooling a portfolio with a new asset forces reduction of the weights of the
extant  assets. In this case also a negative IVaR indicates reduction of
risk. However, we will see below that the set of assets that reduce risk when
pooled with a portfolio is much larger than the set of hedges.

In theory, a straightforward way to calculate IVaR requires us to create a
new portfolio incorporating the candidate new strategy, then reassess VaR
and finally compare it with the previous one. Although in recent time,
techniques for upgrading the construction of the portfolio probability
distributions have been proposed (see Wang [2002]), due to the non-linearity
of VaR, computation may still turn out to be too time-expensive thus being a
bar to real-time decision making. Having at disposal friendly-to-use
approximating formulae may be a first step to overcome this drawback. This
is just the aim of the paper.

Beta-based approximations of IVaR are achieved. Formulae highlight a too
often underestimated aspect. By virtue of the different diversification
impact, conditions for adding new positions in the portfolio are much
stronger than those for their pooling. A final warning. Although the
formulae are more reliable 
for elliptical returns, as we skip behind this assumption
they might grow lower in confidence, since they provide only a linear
approximation. Clearly, higher order approximations should be worked out.
Although this is technically attainable, it drives to cumbersome formulae
which will be in contrast to the spirit of this paper. In conclusion, the
linear approximation formulae can be a useful tool for a preliminary
screening of the alternatives in risk allocation.

The paper is organized as follows. In Section 2 Incremental VaR (IVaR) is
defined. An approximate IVaR formula for risk adding is introduced in
Section 3, whereas one for risk pooling occurs in Section 4. A conclusion in
Section 5 ends the note.

\section{Incremental VaR}

VaR measures the smallest level of under-performance of a position that would
occur with a low probability by a given time horizon. In the following, we
will deal with the \textit{excess return }of a position with reference to a
benchmark. The benchmark may vary with the context: it could be the
liabilities for pension funds, the investment benchmark for traditional
asset managers, or just the cash for hedge funds. Let us give a sketchy
definition list. Define for random a variable $Z$ and $\alpha \in (0,1)$ the
$\alpha $-quantile $q_{\alpha }$ of $Z$ by
\begin{equation}
q_{\alpha }(Z)\quad {\overset{\text{def}}{=}}\quad \inf \{z\in \mathbb{R}%
\,|\,\mathrm{P}[Z\leq z]\geq \alpha \}\,.  \label{eq:2.1}
\end{equation}
In what follows, we use the definition of VaR \emph{relative to the mean} in
the sense\footnote{%
Another possibility would be to use the \emph{absolute} definition $\mathrm{%
VaR}_{\alpha }(Z)\quad {\overset{def}{=}}\quad -q_{\alpha }(Z).$%
\par
Clearly, the two of definitions coincide if $\mathrm{E}(Z)=0.$} of Jorion
[1997, page 87]:\textrm{
\begin{equation}  \label{vardef}
\mathrm{VaR}_{\alpha }(Z)\quad \overset{\text{def}}{=}\quad \mathrm{E}%
(Z)-q_{\alpha }(Z)
\end{equation}
}Let us focus on the impact on the current portfolio by a prospective asset
purchase or disposal. Stand-alone the risk involved in the individual asset,
we need also to take into account the effects of this position on the new
aggregate portfolio. For notation convenience, let us fix $\alpha $ (usually
$\alpha =0.01$ or $\alpha =0.05$) and drop the symbol $\alpha $ in future
formulae. Thus, we define the \textit{Incremental VaR (IVaR) }as the
difference between the VaR of the new and the current portfolio:

\begin{center}
Incremental VaR $=\mathrm{VaR}\left( \text{new portfolio}\right) -\mathrm{VaR%
}\left( \text{current portfolio}\right) $
\end{center}

Clearly, IVaR can be positive, if the candidate strategy adds risk to the
current portfolio, or negative, if the strategy acts like a hedge to the
existing portfolio risks, or zero if it is neutral.

In the sequel, we will deal with two strategies to allocate risk, termed
\emph{risk adding} and \emph{risk pooling}.

\section{IVaR for risk adding}

Let $X$ denote the random excess return of the current portfolio. Suppose we
want to measure the contribution of a position $Y$ or a group of positions
to the total risk of the portfolio. Two situations may occur. We could want
to measure the portfolio risk after having sold $Y$ or a portion $aY$ of it,
and having invested the proceeds in cash. But we could also focus on the
portfolio after having bought the position or a portion $aY$ of it, by
drawing the requested money from cash. In any case, the result is
\begin{equation}
\text{new portfolio }\mathrm{\quad {\overset{def}{=}}\quad }X+aY
\end{equation}
where $a<0$ and $a>0$ refer to the selling and buying case, respectively;
clearly, if $a=0$ no change in the portfolio occurs.

According to the definition of IVaR, we get

\begin{center}
$\mathrm{IVaR} = \mathrm{VaR}\left( X+aY\right) - \mathrm{VaR}\left(
X\right) $
\end{center}

Writing IVaR as a function of the variable $a$, we obtain

\begin{center}
$f\left( a\right) = \mathrm{VaR}\left( X+aY\right) - \mathrm{VaR}\left(
X\right) = a\,\mathrm{E}\left[Y\right] + q(X) - q\left( X+aY\right)$
\end{center}

An easy way to get a rough estimate of IVaR is to look for a linear
approximation. Since $f\left( 0\right) =0$, in case of $f$ being
differentiable with respect to $a$ the sign of IVaR for small positive $a$
is just the sign of $f'(0)$. Conditions for $f$ to be differentiable
and an explicit formula for the derivative are provided in Gouri\'{e}roux at
al. [2000], Lemus [1999], and Tasche [1999]. Application of this formula
yields

\begin{center}
$f^{\prime }\left( a\right) = \mathrm{E}\left[ Y\right] - \mathrm{E}\left[
Y|X+aY=q\left( X+aY\right) \right] $,
\end{center}

where $\mathrm{E}[Y | Z = z]$ denote the conditional expectation of $Y$
given that the random variable $Z$ equals $z$, and in particular

\begin{center}
$f^{\prime }\left( 0\right) = \mathrm{E}\left[ Y\right] - \mathrm{E}\left[
Y|X=q\left( X\right) \right] $.
\end{center}

Observe that $\mathrm{E}\left[ Y|X=q\left( X\right) \right] -$ $\mathrm{E}%
\left[ Y\right] $ is the best prediction of $Y-$ $\mathrm{E}\left[ Y\right] $
given that $X-$ $\mathrm{E}\left[ X\right] =q\left( X\right) -$ $\mathrm{E}%
\left[ X\right] .$ In the case of the conditional expectation $\mathrm{E}%
\left[ Y|X=\cdot\right] $ not being available, a reasonable approximation
would be the best \textit{linear }prediction of $Y-$ $\mathrm{E}\left[ Y%
\right] $ given that $X-$ $\mathrm{E}\left[ X\right] = q\left( X\right) -$ $%
\mathrm{E}\left[ X\right] .$ This linear prediction is given by
\begin{equation}\label{eq:linear}
\left( q\left( X\right) - \mathrm{E}\left[ X\right] \right) \frac{\mathrm{%
cov}\left( X,Y\right) }{\mathrm{var}\left( X\right) },
\end{equation}
where cov and var denote covariance and variance respectively, as usual. By (%
\ref{vardef}), we therefore obtain $f^{\prime }\left( 0\right) \thickapprox
\beta _{yx}\,\mathrm{VaR}\left( X\right) ,$ where $\beta _{yx}=\frac{\mathrm{%
cov}\left( X,Y\right) }{\mathrm{var}\left( X\right) }$ is the standard beta
coefficient of $Y$ with respect to $X.$ This way, a linear approximation of
IVaR comes out as\textrm{
\begin{equation}  \label{ivarapprox}
IVaR \thickapprox \beta _{yx}\,\mathrm{VaR}\left( X\right)\,a.
\end{equation}
}Now, a clear-cut tool for discriminating profitable strategies has emerged.
Just a glance at the sign of $\beta _{yx}$ is sufficient to give a rough
information on the sign of the risk contribution of the position $Y.$

\begin{remark}
\begin{enumerate}
\item  If we are considering the position $aY$ for purchasing (so that $a>0$%
), the negativeness of $\beta _{yx}$ signals that it will act like a risk
diversifier. Vice versa, if we are looking for measuring the risk
contribution of the position $aY$ \ already contained in the portfolio, the
negativeness of $\beta _{yx}$ signals that when sold it will act like a risk
contributor. In conclusion, as intuition suggests, adding $aY$ tends to
reduce the risk only if $Y$ is a super-defensive asset, i.e. the return goes
in the opposite direction of that of $X.$ Vice versa, selling $aY$ tends to
reduce the risk only if $Y$ is a conservative or aggressive asset with
respect to $X,$ i.e. the return of $Y$ goes in the same direction of that of
$X.$

\item  Approximation (\ref{ivarapprox}) holds under very loose conditions on
distributions. For instance, it suffices that $X$ and $Y$ have a continuous
joint density. This is a standard assumption in financial modeling.

\item  Recall that  (\ref{ivarapprox}) comes out as the result of two
  consecutive approximations. Firstly, we have replaced $f(a)$ by the term 
$a f'(0)$.
Secondly, we have substituted (\ref{eq:linear}) for $f'(0)$. It can be proved that the best linear prediction by (\ref{eq:linear}) 
coincides with the best prediction 
$\mathrm{E}\left[ Y|X=q\left( X\right) \right] -$ $\mathrm{E}%
\left[ Y\right] $
if $X$ and $Y$ are normal returns (or more general, if $X$ and $Y$
are jointly elliptical distributed\footnote{%
Elliptical distributions, include the normal distribution as a special case,
as well as the Student's t-distribution and the Cauchy distribution.
Elliptical distributions are called ''elliptical'' since the contours of the
density are ellipsoids. Kelker [1970] proved that VaR can be written in term
of standard deviation when the underlying distribution is elliptical.}\/).
This indicates that approximation (\ref{ivarapprox}) will be reliable for
small $a$ at least in a ``normal'' (or elliptical) world.
Nevertheless, as we skip behind this assumption, the formula has to be
handled with caution since it may lead to a misleading information.

\item One way to
to improve on the approximations 
is working out higher order approximations (see Gouri%
\'{e}roux at al.\ [2000] for the second order derivative of VaR). But a dark
side of the coin exists. The higher the approximation order is the more
cumbersome  the formulas are. So, this approach is just in contrast to the
spirit of this paper.
\end{enumerate}
\end{remark}

\section{IVaR for risk pooling}

As in the previous section, let $X$ be the random excess of return of the
current portfolio. Let $aY$ be the position we are considering to purchase.
The aim in this case is just to arrive at portfolio diversification. If the
asset is purchased, the portfolio is re-balanced, so\textrm{
\begin{equation}
\text{new portfolio }\mathrm{\quad {\overset{def}{=}}\quad }\frac{X+aY}{%
\left( 1+a\right) }
\end{equation}
}where $a>0$ and the factors $\frac{1}{\left( 1+a\right) }$ and $\frac{a}{%
\left( 1+a\right) }$ are the relative weights of the assets $X$ and $Y$,
respectively. Therefore,

\begin{center}
IVaR$= \mathrm{VaR}\left( \text{new re-balanced portfolio}\right) - \mathrm{%
VaR}\left( \text{old portfolio}\right) $.
\end{center}

Proceeding as in the risk adding case, let us write IVaR as a function of
the variable $a$:

\begin{center}
$g\left( a\right) = \mathrm{VaR}\left( \frac{X+aY}{1+a}\right) - \mathrm{VaR}%
\left( X\right) = \mathrm{E}\left[ \frac{X+aY}{1+a}\right] -q\left( \frac{%
X+aY}{1+a}\right) - \mathrm{E}\left[ X\right] +q\left( X\right) $
\end{center}

In particular, we have $g\left( 0\right) =0.$ In case of $g$ being
differentiable with respect to $a$ we obtain

\begin{center}
$g^{\prime }\left( a\right) = {\left( 1+a\right) ^{-2}}\,{\mathrm{E}\left[ Y%
\right] - \mathrm{E}\left[ X\right] + q\left( X+aY\right) -\left( 1+a\right)
\mathrm{E}\left[ Y|X+aY=q\left( X+aY\right) \right] }$,
\end{center}

and in particular
\begin{align*}
g^{\prime }\left( 0\right)& = \mathrm{E}\left[ Y\right] - \mathrm{E}\left[ Y
| X=q\left( X\right) \right] - \mathrm{VaR}\left( X\right) \\
& \approx \left( \mathrm{E}\left[ X\right] - q\left( X\right) \right) {%
\textstyle \frac{\mathrm{cov}\left( X,Y\right) }{\mathrm{var}\left( X\right)
}} - \mathrm{VaR}\left( X\right).
\end{align*}
Therefore $g^{\prime }\left( 0\right) \thickapprox \left( \beta
_{yx}-1\right) VaR\left( X\right) ,$ and the linear approximation of IVaR
turns out to be\textrm{
\begin{equation}  \label{pooling}
\mathrm{IVaR} \thickapprox a \left( \beta _{yx}-1\right) \mathrm{VaR}\left(
X\right).
\end{equation}
}

\begin{remark}
\begin{enumerate}
\item  Although formula (\ref{pooling}) looks similar to (\ref{ivarapprox}),
a striking difference sticks out. In case of pooling, the watershed for
discriminating profitable strategy is no longer the sign of $\beta _{yx},$
but the sign of $\left( \beta _{yx}-1\right) .$ So, it may happen that even
if $\beta _{yx}$ is positive (but less than one) and hence $X$ and $Y$ are
positively correlated, pooling of at least a small portion $aY$ may be
always advisable for reducing the risk. Let us be more precise, consider the
case of $0<\beta _{yx}<1,$ i.e.\ $Y$ is a defensive asset. Because of
positive correlation, $Y$ goes in the same direction as $X$, but $Y$ varies
less than $X$ does. This means that the variations of the excess return $Y$
grow lower than those of $X$ do. In the case of $\beta _{yx}\leq 0,$ i.e.\ $Y
$ is a super-defensive asset, variations of $Y$ go in the opposite direction
of those of $X.$ In conclusion, except in the case of a super-aggressive
asset with $\beta _{yx}>1$, pooling at least a small portion $aY$ is always
an advisable strategy for reducing risk.

\item  The fact that the risk pooling conditions are much looser than those
for risk adding, should not surprise. In a normative framework, the argument
about different attitudes in accepting risk adding and risk pooling can be
tracked back to the so-called ''Samuelson's Fallacy of Large Numbers'' (see
Ross [1999] and the references thereby). By virtue of the favorable
diversification effect, the acceptance of pooling in the portfolio a
sufficiently long string of single-rejected risks is considered
''rational''. Vice versa, the eventual acceptance of adding single-rejected
risks is much more questionable.

\item  Again, (\ref{pooling}) is more reliable 
for elliptically distributed returns,
but may lose its reliability as we relax this assumption.
\end{enumerate}
\end{remark}

\section{Conclusion}

Beta-based approximation formulae for IVaR are achieved in both cases --
adding a new risk and pooling it with the existing portfolio. These
approximations are based on recent results on the differentiability of VaR.
The formulae will give the right indications for trade decisions as long as
the weights of the assets under consideration are small compared to the
weights of the unchanged assets. One can see from the formulae that the
conditions for adding new positions are by far stronger than those for
pooling.

\paragraph{Acknowledgment.} The authors are grateful to Rongwen Wu for
drawing their attention to a misleading formulation in an earlier version of
this paper.

\section*{References}

\begin{description}
\item[Dowd, K. ]  ''Beyond value at risk.'' John Wiley \& Sons, 1998.

\item[Dowd, K. ]  ''A value at risk approach to risk-return analysis.''
\emph{The Journal of Portfolio Management}, 25, Summer 1999, 60--67.

\item[Dowd, K. ]  ''Adjusting for Risk: An Improved Sharpe Ratio.'' \emph{%
International Review of Economics and Finance}, 9, 2000, 209-222.

%\item[Embrechts, P., McNeil, A. and D.Straumann.]  ''Correlation and
%dependency in risk management: properties and pitfalls.'' Preprint, ETH
%Z\"{u}rich, (1998).

\item[Gouri\'{e}roux, C., Laurent, J.~P. and O. Scaillet.]  ''Sensitivity
analysis of Values at Risk.'' \emph{\ Journal of Empirical Finance}, 7,
2000, 225-245.

\item[Kelker, D.]  ''Distribution theory of spherical distributions and a
location-scale parameter generalization.'' \emph{Sankhy\={a}~Ser.~A} 32,
1970, 419--438.

\item[Lemus, G.]  ''Portfolio Optimization with Quantile-based Risk
Measures.'' PhD thesis, Sloan School of Management, MIT. \newline
\texttt{http://citeseer.nj.nec.com/lemus99portfolio.html}

\item[Mina, Jorge. ]  ''Measuring Bets With Relative Value at Risk.'' \emph{%
Derivatives Week}, Learning curve, January 2002, 14-15.

\item[Jorion, P. ]  ''Value at risk: the new benchmark for controlling
market risk.'' Irwin, Chicago 1997.

\item[Ross, S.]  ''Adding Risks: Samuelson's Fallacy of Large Numbers
Revisited.'' \emph{Journal of Financial and Quantitative Analysis}, 34(3),
1999, 323-339.

\item[Tasche, D. ]  ''Risk contributions and performance measurement.''
Working paper, Technische Universit\"{a}t M\"{u}nchen 1999. \texttt{%
http://www.ma.tum.de/stat/}

\item[Wang, Z.]  ''The Properties of Incremental VaR in Monte Carlo
Simulations.''\emph{\ The Journal of Risk Finance}, 3, n. 2, 2002, 14-23.
\end{description}

\end{document}